 \documentstyle[11pt]{article}
 
\newcommand{\be}{\begin{equation}}
\newcommand{\ee}{\end{equation}}
\newcommand{\bea}{\begin{eqnarray}}
\newcommand{\eea}{\end{eqnarray}}
\newcommand{\beann}{\begin{eqnarray*}}
\newcommand{\eeann}{\end{eqnarray*}}
\newcommand{\beasn}{\begin{sneqnarray}}
\newcommand{\eeasn}{\end{sneqnarray}}
\newcommand{\bref}[1]{(\ref{#1})}
\newcommand{\eps}{\epsilon}

\newcommand{\PLB}[3]{{\sl Phys. Lett.} {\bf #1B} (19#2),  {#3}}

\catcode`@=11
\def\dif{{\rm d}}
\def\deriv{\@ifnextchar[{\@deriv}{\@deriv[]}}
   \def\@deriv[#1]#2#3{\mathchoice%
{{\dif^{#1}#2\over\dif{#3}^{#1}}}{{\dif^{#1}#2/\dif{#3}^{#1}}}%
{{\dif^{#1}#2\over\dif{#3}^{#1}}}{{\dif^{#1}#2/\dif{#3}^{#1}}}}

\def\presup#1{{}^{#1}\kern-.15em\relax}      %pre-superscript
\def\presub#1{{}_{#1}\kern-.12em\relax}      %pre-subscript

\def\secteqno{\@addtoreset{equation}{section}%
\def\theequation{\thesection.\arabic{equation}}}
\def\endsecteqno{\def\theequation{\@ifundefined{chapter}%
{\arabic{equation}}{\thechapter.\arabic{equation}}}}
\newcounter{subequation}
\def\thesubequation{\alph{subequation}}
\def\sneqnarray{\stepcounter{equation}\let\@currentlabel=\theequation
\setcounter{subequation}{1}
\def\@eqnnum{{\rm (\theequation\thesubequation)}}
\global\@eqcnt\z@\tabskip\@centering\let\\=\@eqncr\let\@@eqncr=\@@sneqncr
$$\halign to \displaywidth\bgroup\@eqnsel\hskip\@centering
 $\displaystyle\tabskip\z@{##}$&\global\@eqcnt\@ne
 \hskip 2\arraycolsep \hfil${##}$\hfil
 &\global\@eqcnt\tw@ \hskip 2\arraycolsep $\displaystyle\tabskip\z@{##}$\hfil
  \tabskip\@centering&\llap{##}\tabskip\z@\cr}
\def\endsneqnarray{\@@sneqncr\egroup $$\global\@ignoretrue}
\def\@@sneqncr{\let\@tempa\relax
   \ifcase\@eqcnt \def\@tempa{& & &}\or \def\@tempa{& &}
   \else \def\@tempa{&}\fi
     \@tempa \if@eqnsw\@eqnnum\stepcounter{subequation}\fi
     \global\@eqnswtrue\global\@eqcnt\z@\cr}
\def\nobiblabels{\def\@lbibitem[##1]##2{\@bibitem{##2}}}

\catcode`@=12
\def\dddot#1{\hbox{$\mathop{#1}\limits^{\ldots}$}}

\def\W{${\cal W}$}

\def\t{\tau}
\def\e01{\W$^{\rm sp(4)}_{(0,1)}$}
\secteqno

\title{{\bf \W$(2,4)$, Linear and Non-local \W-Algebras
in Sp(4) Particle Model}}

\author{{\sc J. Gomis}$^\dagger$,\ %
        {\sc J. Herrero}$^\dagger$ {\sc and}
        {\sc K. Kamimura}$^\flat$\\
        \llap{$^\dagger$}%
        \small{\it{Departament d'Estructura i Constituents
               de la Mat\`eria}}\\
        \small{\it{Universitat de Barcelona \&}}\\
        \small{\it{Institut de F\'\i sica d'Altes Energies}}\\
        \small{\it{Diagonal, 647}}\\
        \small{\it{E-08028 BARCELONA}}\\
        \llap{$^\flat$}%
        \small{\it{Department of Physics, Toho University}}\\
        \small{\it{Funabashi}}\\
        \small{\it{274 Japan}}\\
        {\it e-mails:} \small{gomis\,@\,rita.ecm.ub.es,
herrero\,@\,ecm.ub.es, KAMIMURA\,@\,JPNYITP}}

\date{}

\begin{document}

\maketitle

\thispagestyle{empty}

\begin{abstract}
We comment on relations between the
linear \W$_{2,4}^{\rm linear}$ algebra and non-linear \W$(2,4)$ algebra
appearing in a Sp(4) particle mechanics model by using Lax equations.
The appearance of the non-local $V_{2,2}$ algebra is also studied.
\end{abstract}

\vfill
\vbox{
\hfill March 1995\null\par
\hfill UB-ECM-PF 95/5\null\par
\hfill TOHO-FP-9550}\null

\clearpage
%%%%%%%%%%%%%%%%%%%%%%%%%%%%%%%%%%%%%%%%%%%%%%%%%%%%%%%%%%%%%%%%%%%%%%%%%
\section{Introduction}
\hspace{\parindent}%

\W-algebras are extensions of the Virasoro algebra \cite{B} with extra
bosonic generators having weights in general greater than 2.
Recently there appeared two interesting generalizations of
the concept of \W-algebra.

\medskip

Linear \W-algebras have been introduced by Krivonos, Sorin and Bellucci
in \cite{linear}.
They have shown that, for certain cases, it is possible to embed a
\W-algebra into a greater non-linear algebra that becomes linear after a
redefinition of its generators.
Several examples are presented explicitly.
For example, by considering a linear algebra spanned by three generators
having weights 1, 2 and 4, a new weight 4 generator
is constructed by means of a non-linear redefinition of the variables
in such a way that the non-linear \W$(2,4)$ algebra becomes
a subalgebra.
Here the additional weight 1 field
plays a crucial r\^ole. Although \W$(2,4)$ is a subalgebra, this
weight 1 field cannot be decoupled from it by a gauge-fixing.

\medskip

The other generalization is the non-local $V_{n,m}$ algebras developed
by Bilal \cite{bilal}.
They are matrix generalizations of \W-algebras which arise in the
context
of non-abelian Toda field theories. The $V_{n,m}$ algebra is the
non-linear symmetry
of a $m$-th order differential equation whose coefficients are $n \times
n$ matrices.

\medskip

In previous papers we have examined a set of particle mechanics models
with Sp$(2M)$ Yang-Mills type gauge symmetry \cite{NOS}. These models
present
various classical \W-symmetries after performing a gauge-fixing induced
by an sl(2) embedding in sp($2M$).

In this paper we will point out that both the \W$_{2,4}^{\rm
linear}$ algebra and
the non-local $V_{2,2}$ algebra are realized at the classical level as
subalgebras of the \W-algebra obtained from the Sp(4) model after
performing the gauge-fixing induced by the sl(2) embedding in sp(4)
with characteristic (0,1) \cite{NOS}. We will refer to this algebra as
\e01. This algebra has four generators:
$H$, $V_2$, $G$ and $C$. The first three generators form a linear
subalgebra which is equivalent to the \W$_{2,4}^{\rm linear}$ examined
in \cite{linear}.
In contrast to ref.\,\cite{linear}, the
weight 1 generator $H$ can be decoupled from the \W$(2,4)$ algebra due
to the presence of the additional weight $0$ generator $C$.
The resulting algebra is equivalent to one obtained by imposing
gauge-fixing conditions on $H$ and $C$.

In \e01 the reduction of the weight 1 generator ($H\,=\,0$)
being $C$ free leads to the non-local algebra $V_{2,2}$. The
non-locality arises from the inverse of the
derivative operator appearing in the definition of the Dirac bracket.
We show it more explicitly by constructing starred variables. They have
non-local forms by the definition. The further reduction of $C$
recovers the locality and the resulting algebra is \W$(2,4)$ again.

\medskip

In the next section we give a brief introduction to the Sp$(4)$ model
and display the classical algebra of the \W-generators.
In section 3 the linear \W-algebra is discussed in the model.
The non-local \W-algebra $V_{2,2}$ is derived in section 4.
A short summary and discussions are in the final section.
A brief sketch of the relations between the algebras is showed in Fig.1.

\unitlength 1mm
\begin{picture}(80,75)(-70,-40)
\put(-6,23){\e01}
\put(-10,13){$(H,V_2,G,C)$}
\put(-6,-20){\W$(2,4)$}
\put(-12,-28){$(H,C)\,\oplus\,(W_2,W_4)$}
\put(-40,0){$V_{2,2}$}
\put(-75,0){$(H)\oplus(V_2^\star,G^\star,C^\star)$}
\put(32,0){\W$_{2,4}^{\rm linear}$}
\put(45,10){$(H,V_2,G\,;\,C)$}
\put(55,6){\vector(0,-1){10}}
\put(45, -10){$(H,V_2,V_4\,;\,C)$}
\put(-35,14){$H \approx 0$}
\put(-35,-15){$C \approx a_0$}
\put(23,14){$C \approx a_0$}
\put(23,-15){$H \approx 0$}
\put(-10,20){\vector(-2,-1){25}}
\put(10,20){\vector(2,-1){25}}
\put(-35, -5){\vector(2,-1){25}}
\put(35,-5){\vector(-2,-1){25}}
\put(-3 ,-37){Fig.\,1}
\end{picture}

%%%%%%%%%%%%%%%%%%%%%%%%%%%%%%%%%%%%%%%%%%%%%%%%%%%%%%%%%%%%%%%%%%%%%

\section{Particle mechanics model with Sp$(4)$ symmetry}
\hspace{\parindent}%

In previous papers \cite{NOS} we have considered a set of models invariant
under a Yang-Mills type gauge symmetry Sp$(2M)$. They are
reparametrization invariant models of $M$ relativistic particles living
in a Minkowskian
$d$-dimensional space-time. For the purpose of the present paper
we give a brief introduction of the Sp$(4)$ case.

The canonical action is given by:
\be
\label{sp can act}
S=\int\dif \t\,\frac12\,\bar R\,{\cal D}\,R,
\ee
where $R$ contains the particle phase space variables and $\bar R$ is
its conjugate:
\be
R\,=\,\left(\begin{array}{c}
        x_1\\x_2\\p_1\\p_2
        \end{array}\right),\quad\quad\quad
\bar R\,=\,\left(\,p_1,\,p_2,-x_1,-x_2 \,\right).
\ee
${\cal D}$ is the covariant derivative with respect to the gauge
group Sp$(4)$:
\be
{\cal D}=\frac{\dif}{\dif \t}-\Lambda,
\quad\quad\quad
\Lambda=\left(\begin{array}{cc}
              B&A\\
             -F&-B^\top
              \end{array}\right)\,\in\, {\rm sp}(4).
\ee
The gauge field $\Lambda$ is a $4 \times 4$ symplectic matrix thus
the components $A$ and $F$ are $2 \times 2$ symmetric matrices.

In this formulation the gauge invariance of the action is expressed in a
manifestly covariant form by means of Yang-Mills type transformations:
\bea
\label{sp mat tra}
\delta R=\beta R,
\eea
\bea
\label{sp gau tra}
\delta\Lambda=\dot\beta-[\Lambda,\beta], \quad\quad\quad
\beta=\left(\begin{array}{cc}
              \beta_B & \beta_A\\
             -\beta_F & -\beta^\top_B
              \end{array}\right)\,\in\, {\rm sp}(4).
\eea
where the gauge parameter $\beta$ is a $4 \times 4$ symplectic matrix.
The equations of motion of the matter fields are
\be
{\cal D}R=\dot R-\Lambda R=0.
\label{mat equ mot}
\ee

The infinitesimal transformation law \bref{sp gau tra} for the gauge
variable $\Lambda$
is the compatibility condition of the pair of equations \bref{sp mat tra} and
\bref{mat equ mot} for the matter variable $R$:
$$
0\,=\,[\,(\delta-\beta),\,{\cal D}\,]\,R\,=\,
-(\delta\Lambda-\dot\beta+[\Lambda,\beta])\,R,
$$
and it can be regarded as a zero-curvature condition. The presence of
a zero-curvature condition allows us to apply the soldering procedure
to reduce the original symmetry of the model to a chiral classical
\W-symmetry by means of a partial gauge-fixing of the $\Lambda$ fields.
For sp$(4)$ we have three different classes of sl$(2)$ embeddings which
will lead to three different gauge-fixings.
For the purpose of the present paper we will examine one of these
embeddings, namely the sl$(2)$ embedding with characteristic $(0,1)$
\cite{NOS}. This embedding induces a gauge-fixing which leaves four
remnant gauge fields: $H$, $T$, $G$ and $C$. After the gauge-fixing
the gauge field $\Lambda$ is given by:
\be
\label{cccc}
\Lambda_r=\left(\begin{array}{cccc}
                H  &  0  &  0  &  1  \\
                0  & -H  &  1  &  0  \\
                C  &  \frac{T}{2}  & -H  &  0  \\
                \frac{T}{2}  &  G  &  0  &  H
              \end{array}\right).
\ee
In this gauge the action \bref{sp can act} becomes,
after integrating over the momenta,
\be
\label{swfo}
S=\int\dif \t\left[(\dot x_1-H x_1)(\dot x_2+H x_2)+
\frac12\left(C x_1^2+Tx_1x_2+Gx_2^2\right)
\right].
\ee
The equations of motion for the matter variables become
\be
\label{eq01}
\left(\begin{array}{cc} C & -(\frac\dif{\dif \t} + H)^2 + \frac12 T \\
                  -(\frac\dif{\dif \t} - H)^2 + \frac12 T & G
\end{array}\right)
\left(\begin{array}{c} x_1 \\ x_2  \end{array}\right)\,=\,0.
\ee
They can be regarded as the Drinfel'd-Sokolov equations for this
embedding.
The corresponding Lax operator is given in a $2 \times 2$ matrix form.

\vskip 3mm

This system has four residual gauge symmetries associated with the
four remnant gauge fields. They are:

\vskip 3mm

$\bullet \eps$-transformation (Diffeomorphisms):
\bea
\nonumber
&\delta H=\eps\dot H+H\dot\eps+\frac{k}{2}\ddot\eps,\hskip 7mm
\delta T=\eps\dot T+2\dot\eps T-\dddot\eps,
\\\nonumber
&\delta C=\eps\dot C+(2-k)C\dot\eps, \hskip 7mm
\delta G=\eps\dot G+(2+k)G\dot\eps,
\\\nonumber
&\delta x_1=\eps\dot x_1+\frac12(k-1)x_1\dot\eps, \hskip 7mm
\delta x_2=\eps\dot x_2-\frac12(k+1)x_2\dot\eps,
\eea
where $k$ is an arbitrary constant coming from the fact that $H$ is
an element of Cartan subalgebra.
The matter variables $x_1$ and $x_2$
transform as primary fields under diffeomorphisms with weights
$\frac12(k-1)$ and $-\frac12(k+1)$ respectively.
The gauge variables $C$ and $G$ transform also as primary fields with
weights $2-k$ and $2+k$.
On the other hand $T$ is a quasi-primary field with
weight $\,2\,$ and $H$ transforms as a field of weight $\,1\,$ with a
$\ddot \epsilon$ term.

\vskip 4mm

$\bullet \alpha$-transformation (Dilatations):
$$
\delta H=\frac12 \dot\alpha,\quad\quad\delta T=0,\quad\quad
\delta C=-\alpha C,\quad\quad\delta G=\alpha G,
$$
\be
\delta x_1=\frac12 \alpha x_1,\quad\quad\delta x_2=-\frac12 \alpha
x_2,
\ee

\vskip 4mm

$\bullet \beta_2$-transformation:
$$
\delta H=\frac12C\beta_2,\quad\delta T=\beta_2(\dot
C-2CH)+2\dot\beta_2C, \quad\delta C=0,
$$
$$\delta G=\beta_2(4H^3-2HT-6H\dot H+\frac12\dot T+\ddot H)
-\dot\beta_2(6H^2-T-3 \dot H)+3H\ddot\beta_2-\frac12\dddot\beta_2,
$$
\be\delta x_1=\beta_2(2Hx_2+\dot x_2)-\frac12x_2\dot\beta_2
\quad\quad\delta x_2=0,
\ee

\vskip 4mm

$\bullet \beta_5$-transformation:
The $\beta_5$ transformations can be obtained from
the $\beta_2$ transformations by the following replacements:
\be
\label{b2b5 cha}
\beta_2\leftrightarrow\beta_5,\quad\quad
H\leftrightarrow -H,          \quad\quad
C\leftrightarrow G,           \quad\quad
x_1\leftrightarrow x_2.
\ee

\medskip
%%%%%%%%%%%%%%%%%%%%%%%%%%%%%%%%%%%%%%%%%%%%%%%%%%%%%%%%%%%

The generators of these transformations are
$H$ for the dilatations, $C$ and $G$ for the $\beta_2$ and $\beta_5$
transformations, respectively, and
\be
V_2\, =\, -\, T\, -\,2\, H^2\,+\,2k\, \dot H
\label{fdpu}
\ee
for the diffeomorphisms. The Poisson brackets of these generators
are\footnote{All fields on the right hand side of the Poisson brackets
depend on $\t$, the dot means derivative with respect to $\t$ and
$\delta \equiv \delta(\t-\t')$}:
\bea
\label{pbHH}
&\{\,H(\t),\,H(\t')\,\} \, = \, -\frac14\,\dot\delta,
\\
&\{\,H(\t),\,V_2(\t')\,\} \, = \, H\,\dot\delta\,+\,\dot
H\,\delta\,+\,\frac{k}{2} \,\ddot\delta,
\\
&\{\,H(\t),\,G(\t')\,\} \, = \, \frac12\,G\,\delta,
\\
&\{\,H(\t),\,C(\t')\,\} \, = \, -\,\frac12\,C\,\delta,
\\
&\{\,V_2(\t),\,V_2(\t')\,\} \, = \, 2 \, V_2\,\dot\delta\,+\,\dot
V_2 \, \delta\,+\, (1+k^2)\,\dddot\delta,
\\
&\{\,G(\t),\,V_2(\t')\,\} \, = \, (2+k)\, G\,\dot\delta\,+\,\dot G
\,\delta,
\\
&\{\,C(\t),\,V_2(\t')\,\} \, = \, (2-k)\, C\,\dot\delta\,+\,\dot C
\,\delta,
\\
&\{\,C(\t),\,C(\t')\,\} \, = \, \{\,G(\t),\,G(\t')\,\}\,=\,0
\eea
and
\bea
\nonumber
\{\,G(\t),\,C(\t')\,\} & = & \left(-8\,H^3\,-\,2\,H\,V_2\,+
\,(4k+8)\,H\,\dot H\,-\,(1+k)\,\ddot H\,+\, \frac12 \,\dot
V_2 \right) \,\delta\,+
\\
\label{pbcg}
&& + \,\left( 8\,H^2\,-\,(3+2k)\,\dot H\,+\,V_2 \right)\,
\dot\delta\,-\,3\,H \,\ddot\delta\,+\,\frac12\,\dddot\delta.
\eea

The Poisson brackets (PB)  \bref{pbHH}--\bref{pbcg}
between the generators $H$, $V_2$, $G$ and $C$
are linear except for the last one $\{G,C\}$.
Furthermore $H$, $V_2$ and $G$ form a subalgebra. This linear subalgebra
happens to be equivalent (for $k=2$) to the \W$^{\rm linear}_{2,4}$
algebra of ref.\,\cite{linear}.

In the next section we are going to show the
relations between \e01, \W$^{\rm linear}_{2,4}$ and \W$(2,4)$
from the study of the Lax equations.

\medskip

%%%%%%%%%%%%%%%%%%%%%%%%%%%%%%%%%%%%%%%%%%%%%%%%%
\section{Linear \W-algebra from Lax equation}
\hspace{\parindent}%

Let us consider the equations of motion \bref{eq01}
for the matter variables
or Lax equations for the \e01 algebra.
Assuming the division by $C$ we can obtain a fourth
order equation for $x_2$,
\be
x_2^{(4)}\,+\,u_1\,\dddot x_2\,+\,u_2\,\ddot x_2\,+\,u_3\,\dot
x_2\,+\,u_4\,x_2\,=\,0.
\ee
The coefficient of the third order derivative term is
$\,u_1=-2\frac{\dot C}{C}$.
It disappears if the equation is expressed for
$\,y=C^{-1/2}x_2$:
\be
\label{lax4}
y^{(4)}\,+\,W_2\, \ddot y\,+\,\dot W_2\, \dot y\, +\,
\left( W_4\,+\,{\frac{9}{100}} \, W_2^2\,+\,{\frac{3}{10}}\,\ddot
W_2 \right)\,y\,=0,
\ee
where $W_2$, $W_4$ are given by:
\be
\label{w2}
W_2\,=\,-\,T\,-\,2\,h^2 \, + \, 4\,\dot h,\,\,\,\,\,\,\,\,\,\,
h\,\equiv\,H+{\dot C \over{2C}}
\ee
and
\bea
\nonumber
W_4 & = &
-\,C\,G\, + \, \frac{4}{25} \, W_2^2 \, + \, \frac15 \, \ddot W_2 \,
+ \, 2 \, h^2 \, W_2 \, - \, 2\,\dot h\,W_2 \, - \,h \, \dot W_2 \, +
\\
&& + \, 4\,h^4 \, + \, 7\,{\dot h}^2 \, - \, 16\,h^2\,\dot h\, + \,
6\,h\,\ddot h \, - \, \dddot h.
\label{w4}
\eea
Note that $h$ is a field having zero PB with $H$.
The Poisson brackets between $W_2$ and $W_4$ close giving the
classical \W$(2,4)$ algebra:
\bea
\label{pbv2v2}
& \{\,W_2(\t),\,W_2(\t')\,\} \, = \, 2\, W_2\,\dot\delta\,+\,\dot
W_2\,\delta\,+\, 5\,\dddot\delta,
\\
\label{pbv2v4}
& \{\,W_4(\t),\,W_2(\t')\,\} \, = \, 4\, W_4\,\dot\delta\,+\,\dot
W_4\,\delta,
\\\nonumber
& \{\,W_4(\t),\,W_4(\t')\,\} \, = \,
     \frac{1}{20} \, \delta^{(7)} \, + \,
     \frac{7}{25} \, W_2 \, \delta^{(5)} \, + \,
     \frac{7}{10} \, \dot W_2 \, \delta^{(4)} \, +
\\\nonumber
& + \, \left(
     \frac{21}{25} \, \ddot W_2 \, + \,
     \frac{49}{125} \, W_2^2 \, + \,
     \frac35 \, W_4
   \right) \, \dddot\delta \, + \,
 \left(
     \frac{14}{25} \, \dddot W_2 \, + \,
     \frac{147}{125} \, W_2\,\dot W_2 \, + \,
     \frac{9}{10} \, \dot W_4
   \right) \, \ddot \delta \, +
\\\nonumber
& + \, \left(
     \frac15 \, W_2^{(4)} \, + \,
     \frac{88}{125} \,W_2 \, \ddot W_2 \, + \,
     \frac{59}{100} \, \dot W_2^2 \, + \,
     \frac{72}{625} \, W_2^3 \, + \,
     \frac12 \, \ddot W_4 \, + \,
     \frac{14}{25} \, W_2\, W_4
    \right) \, \dot\delta \, +
\\\nonumber
& +  \left(
     \frac{3}{100}\, W_2^{(5)} \, + \,
     \frac{177}{500}\, \dot W_2\,\ddot W_2 \, + \,
     \frac{39}{250} \, W_2\, \dddot W_2 \, + \,
     \frac{108}{625}\, W_2^2 \, \dot W_2 \, + \,
     \frac{1}{10} \, \dddot W_4 \, + \,
     \frac{7}{25} \, (W_4 W_2)^{\dot{\ }}
   \right)  \delta.
\\
& \label{pbv4v4}
\eea
Furthermore $W_2$ and $W_4$ have zero Poisson bracket both with $H$ and
$C$. This implies that $W_2$ and $W_4$ are combinations invariant under
$\alpha$ and $\beta_2$ transformations.
Thus we can gauge fix  $H$ and $C$ independently of $W_2$ and $W_4$.
This decoupling procedure of $C$ and $H$ is equivalent to imposing
a set of gauge fixing conditions: $H=0$ and $C=a_0$, constant. Regarding
them as second class constraints and eliminating them we obtain the
non-linear \W$(2,4)$ algebra realized by Dirac brackets.

The appearance of the \W$(2,4)$ algebra is a consequence of the
transformation of the \e01 Lax equations \bref{eq01} into the
\W$(2,4)$ Lax equation \bref{lax4}. The procedure described above gives
in an easy way the expression of the \W$(2,4)$ generators in terms of
the \e01 generators \bref{w2}--\bref{w4}.

\medskip

If we further analyze the algebra ($H$, $W_2$, $W_4$, $C$) by imposing
the
gauge-fixing condition $C=a_0$, where $a_0$ is a non-vanishing constant,
we will find out the appearance of a linear algebra.
Since we are interested in the gauge fixing $C=a_0$, the conformal
weight of $C$ should be zero.
Therefore we should change the energy momentum tensor.
The corresponding energy-momentum tensor
is given by $V_2$ \bref{fdpu} with $k=2$:
\be
\label{V2T}
V_2\,=\,-\,T\,-\,2\,H^2 \, + \, 4\,\dot H,
\ee
which is obtained from $W_2$ \bref{w2} by putting $C=a_0$, {\it i.e.}
by replacing $h$ by $H$.
The new weight $4$ generator, $V_4$, is also given by the same
replacement in $W_4$:
\bea
\nonumber
V_4 & = &
-\,a_0\,G\, + \, \frac{4}{25} \, V_2^2 \, + \, \frac15 \, \ddot V_2 \,
+ \, 2 \, H^2 \, V_2 \, - \, 2\,\dot H\,V_2 \, - \,H \, \dot V_2 \, +
\\
\label{defV4}
&& + \, 4\,H^4 \, + \, 7\,{\dot H}^2 \, - \, 16\,H^2\,\dot H\, + \,
6\,H\,\ddot H \, - \, \dddot H.
\eea
The fields $V_2$ and $V_4$ close under Poisson brackets forming again
the non-linear \W$(2,4)$ algebra. The interesting observation is that
the generators $H$, $V_2$ and $G$ form a linear algebra, which is
equivalent to the \W$^{\rm linear}_{2,4}$ algebra of ref.\,\cite{linear}
and, therefore, equation \bref{defV4} gives in a natural way the
non-linear change of variables that relates the linear algebra
($H$, $V_2$, $G$) with the non-linear one ($V_2$, $V_4$).

The relations between the different algebras can be summarized in this
scheme:
\bea
\nonumber
{\cal W}^{\rm sp(4)}_{(0,1)} = (H, V_2, G, C)
& \supset & {\cal W}_{2,4}^{\rm linear} = (H, V_2, G)
\\\nonumber
\downarrow \quad\quad\quad & & \quad\quad\quad \downarrow
\\\nonumber
{\cal W}(2,4) \oplus (H,C) & & (H, V_2, V_4) \supset {\cal
W}(2,4)
\eea
The vertical arrows symbolize a non-linear change of variables induced by
the Lax equations.

We believe that the study of the Lax equations described above could be
useful to construct other classical linear \W-algebras and the non-linear
changes relating them to their non-linear counterparts.

\medskip

%%%%%%%%%%%%%%%%%%%%%%%%%%%%%%%%%%%%%%%%%%%%%%%%%%%%%%%%%%%%%%%%%%%%%%%

\section{Non-local \W-algebra}
\hspace{\parindent}%

Non-local extensions of \W-algebras were discussed by Bilal
\cite{bilal}. We can see how the simplest one, $V_{2,2}$, arises in our
model.
It appears after a reduction of $H$ in the \e01 algebra of
section 2.
This reduction is possible by defining the Dirac bracket or by
introducing starred variables. We follow the latter approach since it
will make clear the reason why the non-linear algebra appears.

Starred variables have zero PB with second-class
constraints $\phi^a$. Their primitive form is:
\be
\label{star}
A^\star\,=\,A\,-\,\{A,\phi^a\}(\{\phi^a,\phi^b\})^{-1}\,\phi^b,
\ee
which is valid up to linear terms of constraints, {\it i.e.},
terms containing higher powers of constraints are not taken into
account.
Here we take $H(\t)$ as $\phi^a$ and keep all orders of constraints.
$H^\star$ is zero by the definition \bref{star}. Others are:
\be
V_2^\star\,=\,V_2+2H^2-2k \dot H\,=\,-\,T
\ee
and
\be
G^\star(\t)\,=\,G(\t)\,e^{-2\int^\t d\t'H(\t')}, \quad\quad\quad
C^\star(\t)\,=\,C(\t)\,e^{2\int^\t d\t'H(\t')}.
\ee
They have zero PB with $H$:
\be
\{V_2^\star,H\}\,=\,\{G^\star,H\}\,=\,\{C^\star,H\}\,=\,0.
\ee
The origin of the non-locality is the inverse of $\,\dot\delta\,$ in the
PB $\{H(\t),H(\t')\}$ \bref{pbHH}.

The three starred generators ($V_2^\star$, $G^\star$ and $C^\star$)
span the non-local algebra $V_{2,2}$ discussed by Bilal \cite{bilal}
under Poisson brackets:
\bea
&\{V_2^\star(\t),V_2^\star(\t')\} \, = \,
2\,V_2^\star(\t)\,\dot\delta\,+\,
\dot V_2^\star(\t)\,\delta\,+\, \dddot\delta,
\\
&\{C^\star(\t),V_2^\star(\t')\} \, = \,
2\,C^\star(\t)\,\dot\delta\,+\,\dot C^\star(\t)\, \delta,
\\
&\{G^\star(\t),V_2^\star(\t')\} \, = \, 2\,G^\star(\t)\,\dot\delta \,+\,
\dot G^\star(\t)\,\delta,
\\
&\{C^\star(\t),C^\star(\t')\} \, = \,
\frac12\,C^\star(\t)\,C^\star(\t')\,\epsilon(\t'-\t),
\\
&\{G^\star(\t),G^\star(\t')\} \, = \,
\frac12\,G^\star(\t)\,G^\star(\t')\,\epsilon(\t'-\t),
\\
&\{C^\star(\t),G^\star(\t')\}  =
-\frac12\,C^\star(\t)\,G^\star(\t')\,\epsilon(\t'-\t)\,+
\,V_2^\star(\t)\,\dot\delta\,+\,\frac12\,\dot V_2^\star(\t)\,\delta +
\frac12\dddot\delta.
\eea

This result is also expected from the fact that the equation of motion
of the model
\bref{eq01} is a $2 \times 2$ matrix second-order Lax equation
and the
condition $H=0$ guarantees the absence of first order derivative term.
$C^{\star}$ and $G^{\star}$ correspond to $V^{\pm}$ of
ref.\,\cite{bilal}.
$C^{\star}$ and $G^{\star}$ are real generators while $V^{+}$ and $V^{-}$
appear as mutually conjugate ones.
We can further reduce $C^\star$ by constructing generators
$V_2^{\star\star}$ and $V_4^{\star\star}$ having zero PB both with $H$
and $C$. Generators satisfying this requirement have been found
and are precisely $W_2$ in \bref{w2} and $W_4$ in \bref{w4}. The
non-locality disappeared
since $G^\star$ and $C^\star$ appear by their product in $W_4$.
The recovery of locality is understood by the fact that the matrix
of constraints,
\be
\pmatrix{\{H(\t),\,H(\t')\}\,&\,\{H(\t),\,C(\t')\}\cr
         \{C(\t),\,H(\t')\}\,&\,\{C(\t),\,C(\t')\}}
\,=\,\pmatrix{-\frac{1}{4}\,\partial_{\t}\delta&-\frac{C}{2}\,\delta\,\cr
\,\frac{C}{2}\,\delta& 0}
\label{pbHHCC}
\ee
has a local inverse:
\be
\pmatrix{0\,&\,\frac{2}{C}\,\delta\cr
         -\,\frac{2}{C}
\,\delta\,&\,-\,\frac{1}{C}\,\partial_{\t}\, (\frac{1}{C}\,\delta)}.
\ee

\medskip

%%%%%%%%%%%%%%%%%%%%%%%%%%%%%%%%%%%%%%%%%%%%%%%%%%%%%%%%%%%%%%%%%%%%%%%%%%%

\section{Summary}
\hspace{\parindent}%

In this paper we have shown that \e01, which is the symmetry algebra of
the Sp$(4)$ particle mechanics model gauge-fixed by means of the $(0,1)$
sl$(2)$ embedding,
encodes both linear and non-local \W-algebras. We reduce the fields $H$
and $C$ one by one. By first reducing
$H$ we get the non-local $V_{2,2}$ algebra \cite{bilal}. On the other
hand, the reduction of $C$ first gives a linear \W-algebra
\cite{linear}.
Reduction of both $H$ and $C$ ends with \W$(2,4)$ any way.

The imposition of $H=0$ is regarded as a gauge-fixing condition by
itself. The non-locality enters from the inverse of the Poisson bracket
\bref{pbHH} appearing in the definition of Dirac bracket.
Furthermore, by imposing $C=a_0$ the non-locality disappears due to the
locality of the inverse of the PB matrix \bref{pbHHCC} and the
resulting residual algebra is \W$(2,4)$.

On the other hand, reduction of $C$ in \e01
leaves three generators ($H$, $V_2$ and $G$) satisfying a linear
algebra which is equivalent to the linear \W-algebra examined in
ref.\,\cite{linear}.
\W$(2,4)$ generators are constructed through a non-linear and
invertible transformation \bref{defV4}.
In contrast to the previous case ($H=0$), the reduction $C=a_0$
is not regarded as a gauge-fixing condition since
$\{C,C\}=0$
does not have inverse --- it is a first class constraint and is regarded
as a restriction on the solutions.
The algebra generated by $H$, $V_2$ and $G$ is a subalgebra of the
original \e01 algebra but not commuting with $C$.
Finally \e01 is decomposed into two mutually decoupled algebras
with generators $(H,C)$ and $(W_2,W_4)$. The latter are given in
\bref{w2} and \bref{w4} and are generators of \W$(2,4)$, which
is the residual symmetry after reducing both $H$ and $C$.

We emphasize that the linear algebra and the non-linear change of
variables that relates it with the non-linear \W-algebra appear
naturally from the study of the Lax equations.

The generalization of present discussions to other cases is interesting.
By considering Sp$(6)$ models \cite{NOS} we may discuss the relations
between the linear \W$^{\rm linear}_3$, non-local $V_{3,3}$ and non-linear
\W$_3$ and \W$_3^2$ algebras.
Finally we add a remark about two papers on linear \W-algebras that have
appeared recently \cite{madsen}. Their results coincide with our
conclusions when they overlap.

\vspace{10mm}

{\bf Acknowledgements:}
This work has been partially supported by CICYT (contract
AEN93-0695) and by the Commission of the European Communities
(contract CHRX-CT93-0362(04)).
J.H. acknowledges a fellowship from Generalitat de Catalunya.

%%%%%%%%%%%%%%%%%%%%%%%%%%%%%%%%%%%%%%%%%%%%%%%%%%%%%%%%%%%%%%%%

\end{document}